\documentclass[12pt,a4paper,fleqn]{article}
\usepackage[T1]{fontenc}
\usepackage{amsmath}
\usepackage{amssymb} 
\usepackage[small]{caption2} 
\usepackage{graphicx} 
\usepackage{mathrsfs}	
\usepackage[small,loose]{subfigure}  
\usepackage{cite} 
\usepackage{collref} 
\usepackage{cancel}
\usepackage{psfrag} 
\usepackage{color} 
\usepackage[english]{babel}
\usepackage{sidecap}
\newcommand{\eVdist}{\kern-0.06em}

\newcommand{\gev}{\:\text{Ge\eVdist V}}
\newcommand{\tev}{\:\text{Te\eVdist V}}

\newcommand{\cm}{\:\text{cm}}

\newcommand{\km}{\:\text{km}}

\newcommand{\s}{\:\text{s}}

\DeclareMathAlphabet{\mathpzc}{OT1}{pzc}{m}{it}

\allowdisplaybreaks

\begin{document}

\begin{titlepage}

\begin{flushright}
DESY 12-230
\end{flushright}

\vspace*{1.0cm}

\begin{center}
{\LARGE\bf
The heterotic MiniLandscape and the\\[0.1cm] 126 GeV Higgs boson
}

\vspace{1cm}

\textbf{
Marcin Badziak\footnote[1]{Email: \texttt{mbadziak@fuw.edu.pl}}$^a$,
Sven Krippendorf\footnote[2]{Email: \texttt{krippendorf@th.physik.uni-bonn.de}}{}$^b$,
Hans Peter Nilles\footnote[3]{Email: \texttt{nilles@th.physik.uni-bonn.de}}{}$^b$,
Martin Wolfgang Winkler\footnote[4]{Email: \texttt{martin.winkler@desy.de}}{}$^c$
}
\\[5mm]
\textit{\small
{}$^a$ Institute of Theoretical Physics, Faculty of Physics, University of Warsaw,\\
ul.\ Ho\.za 69, PL--00--681 Warsaw, Poland.}
\\[3mm]
\textit{\small
{}$^b$ Bethe Center for Theoretical Physics {\footnotesize and} 
Physikalisches Institut der Universit\"at Bonn,\\
Nussallee 12, 53115 Bonn, Germany.
}
\\[3mm]
\textit{\small
{}$^c$ Deutsches Elektronen-Synchrotron DESY, Notkestrasse 85, 22607 Hamburg, Germany.
}
\end{center}

\vspace{1cm}

\begin{abstract}
The MSSM candidates arising from the heterotic MiniLandscape feature a very constrained supersymmetry breaking pattern. This includes a fully predictable gaugino mass pattern, which is compressed compared to the CMSSM, and an inverted sfermion hierarchy due to distinct geometric localisation, featuring stops as light as $ 1\tev$. The observed Higgs mass sets a lower bound $m_{\widetilde{g}}>1.2 \tev$ on the gluino mass. The electroweak fine-tuning is reduced by a UV relation between the scalar mass of the two heavy families and the gluino mass. While large parts of the favoured parameter space escape detection at the LHC, the prospects to test the MiniLandscape models with future dark matter searches are very promising.

\end{abstract}

\end{titlepage}

\newpage
\tableofcontents

\section{Introduction}

In the context of heterotic orbifold compactifications a large class of MSSM models has been constructed, known as the MiniLandscape~\cite{Lebedev:2006kn,Lebedev:2008un,Lebedev:2007hv}. In this MiniLandscape supersymmetry is broken by a mixture of moduli mediation and anomaly mediation, referred to as Mirage scheme~\cite{0411066,0503216,0504036, Lowen:2008fm}. As discussed in~\cite{Krippendorf:2012ir}, the moduli mediated contribution to scalar masses depends on the localisation of matter fields in the extra dimensions. There are two distinct classes of matter fields, namely matter fields arising in the untwisted sector and fields in twisted sectors. The Higgs fields and the top quark arise in the untwisted sector whereas the other fields generally arise in the twisted sector. The top quark is located in the untwisted sector to generate a large top quark Yukawa coupling. This results in soft scalar masses for untwisted fields that are suppressed compared to scalar masses for twisted fields, a UV realisation of the scheme known as {\it natural SUSY.}

Furthermore, the breaking to the Standard Model gauge group by turning on appropriate Wilson lines fixes the rank of the hidden sector gaugino condensate. In a significant fraction (more than 70 percent) of MiniLandscape models it leads to a gravitino mass in the multi-TeV range, i.e. low-energy SUSY is realised without additional fine-tuning of the gravitino mass~\cite{Lebedev:2006tr}. This breaking also fixes the ratio between anomaly mediated and moduli mediated supersymmetry breaking, leading to a clear phenomenological relation between gravitino and gaugino masses, allowing a prediction for the ratio among the gaugino masses.

The resulting pattern of soft-masses has only very few parameters, the gravitino mass $m_{3/2}$ (within the multi-TeV range), the twisted and untwisted sector scalar masses $m_1$ and $m_3$ (which should satisfy $m_1\gg m_3$) as well as the $\mu$ and $B\mu$ terms. A special field theoretical engineering of the MSSM twisted sector fields could in principle decouple their mass from the gravitino mass, but we restrict ourselves to the case $m_1\leq m_{3/2}$, i.e. no enhancement compared to the scale of supersymmetry breaking. 

Given the recent experimental results on supersymmetry~\cite{:2012mfa,CMS-PAS-SUS-12-028,ATLAS-CONF-2012-109,ATLAS-CONF-2012-145} and in particular the Higgs results~\cite{atlashiggs,cmshiggs}, the goal of this paper is to identify the distinct phenomenological features of this constrained model which can be tested by current LHC and dark matter experiments. In particular we identify the following properties:
 \begin{itemize}
\item Depending on the rank of the condensing hidden sector gauge group, the ratio among the gaugino masses is fixed to be
\begin{equation}
(m_{\widetilde{B}}:m_{\widetilde{W}}:m_{\widetilde{g}})~=~ (1:1.3:2.6) \;\text{ or }\; (1:1.4:2.9)\;.
\end{equation}
\item A Higgs mass $m_h=125-126\gev$ limits the possible gravitino mass to $m_{3/2}>15\tev$. This translates into a lower bound on the gluino mass which is given by $1.2\tev$.\footnote{This bound can in principle be lowered depending on the theoretical uncertainty for the Higgs mass.}
 \item Parts of the parameter space lead to a realisation of suppressed third generation squark masses, in particular the stops can be as light as $\sim 1\tev,$ while $m_1$ is of order $m_{3/2}.$ In this range a sufficiently large Higgs mass can be realised through large stop mixing which is generated radiatively similar as in~\cite{Badziak:2012rf}.

\item  Dark matter can be a bino-like or a higgsino-like neutralino where the correct relic density can be obtained by stop coannihilations (bino case) or non-thermal production (higgsino-case). The composition of the lightest neutralino is influenced by the geometric location of $\tau_R$ in the twisted or untwisted sector. In particular we find the lightest neutralino to be higgsino-like mostly in models with $\tau_R$ in the untwisted sector and bino-like for models with $\tau_R$ in the twisted sector.

\end{itemize}
A light higgsino for models with $\tau_R$ in the untwisted sector, arises due to the UV relation between the masses of the gauginos and the scalars of the twisted sector. This relation results in a significant cancellation among radiative contributions to the soft mass of the up-type Higgs $m_{H_u}$. This, in terms, leads to a suppressed $|\mu|$ (higgsino mass) which is related to $m_{H_u}$ through electroweak symmetry breaking conditions. Intriguingly, a direct consequence of this effect is a reduced electroweak fine-tuning.

The rest of this article is structured as follows. We first review the UV structure of SUSY breaking in the heterotic MiniLandscape, highlighting the relation between the constraints on hidden sector gaugino condensation and the appearing suppression of soft masses (section~\ref{sec:topdownnatural}), i.e.~fix the relation between gravitino and gaugino masses. In section~\ref{sec:phenoanalysis} we then analyse the phenomenological properties of this class in detail before concluding in section~\ref{sec:conclusions}.

\section{UV SUSY breaking in the MiniLandscape}
\label{sec:topdownnatural}
In models of the heterotic MiniLandscape SUSY is broken in the process of moduli stabilisation, leading to a scenario with mixed moduli and anomaly mediated SUSY breaking. In~\cite{Lowen:2008fm} it was shown that given a supersymmetric stabilisation of K\"ahler and complex structure moduli, a mirage pattern of suppressed gaugino masses and A-parameters compared to the gravitino mass (originally found in the context of type IIB KKLT compactifications~\cite{0411066,0503216,0504036}) also arises in models of the heterotic MiniLandscape. This mirage pattern for the gaugino masses is augmented by distinct scalar masses for MSSM matter fields arising from the twisted and untwisted sector~\cite{Krippendorf:2012ir}. In particular the geometric localization of the third generation can lead to a UV realization of the bottom-up scheme known as {\it natural SUSY.} Here we would like to discuss briefly how the suppression of the gauginos is fixed in this UV scheme and then shortly summarize the soft-mass pattern with its free parameters.

\subsection{The SUSY breaking scales in Mirage mediation}
The suppression of gaugino masses (and A-parameters) is determined in the UV scheme as follows (see~\cite{0411066,0503216,0504036} for the corresponding discussion in the type IIB KKLT setup). After stabilising the K\"ahler and complex structure moduli supersymmetrically~\cite{Kappl:2010yu,Anderson:2011cza}, we are left with an effective four-dimensional theory for the yet unfixed dilaton $S,$ which should be of the following form to achieve a de Sitter stabilisation of the dilaton
\begin{eqnarray}
K&=& -\log{(S+\bar{S})}+K_{\rm up}(X,\bar{X})\;,\\
W&=& C+ P e^{-bS}+ W_{\rm up}(X)\;,
\end{eqnarray}
where $P$ and $C$ denote constants, $b$ is linked to the rank of the condensing hidden sector gauge group (for SU(N): $b=8\pi^2/N$), and $K_{\rm up}(X,\bar{X})$ and $W_{\rm up}(X)$ specify the K\"ahler and superpotential of the hidden sector matter uplifting sector. For an explicit example of such an uplifting sector along with more details on the appearance/construction of the dS minimum we refer the reader to appendix~\ref{sec:detrho}. The gravitino mass is set by the gaugino condensate
\begin{equation}
m_{3/2}=\frac{|W|}{\sqrt{2s}}=\sqrt{2} P b e^{- b s_0} s_0=\frac{16\pi^2\, P}{N}\  e^{-\frac{16\pi^2}{N}}\, ,
\end{equation}
where we used in the last step that the dilaton is stabilised at $s_0=2$ which is required to correctly reproduce the value of the unified gauge coupling at the high scale. Assuming $P$ to be ${\mathcal O}(1),$ the rank of the gaugino condensate sets the overall scale of the soft parameters. As discussed in~\cite{Lebedev:2006tr}, the models in the MiniLandscape (more than 70 percent) feature a hidden sector gaugino condensate with rank $N=4,5$ and hence lead to realisation of low-energy supersymmetry without any additional fine-tuning.

In this approach to moduli stabilisation, supersymmetry is predominantly broken by the hidden uplifting sector $F_X\neq 0$ and the dilaton F-term $F_S$ is only non-vanishing at sub-leading order. Explicitly one finds
\begin{eqnarray}
F_X&\simeq&\sqrt{3}m_{3/2}=2 \sqrt{3}  P b e^{-2b}\, ,\\
F_S &=& e^{K/2}K^{SS}D_SW\simeq 6P e^{-2b}=\frac{3 m_{3/2}}{b}\, .
\end{eqnarray}

\subsection{The scales of soft-masses in the MiniLandscape}
\label{sec:scalesinminilandscape}
{\bf Gaugino masses:} This difference in the F-terms becomes important when looking at the soft scalar masses, in particular at the gaugino masses and A-terms which are not generated by $F_X$ but by $F_S.$ For instance, the moduli mediated gaugino masses are found to be
\begin{equation}
M_a=\frac{F_S}{2 s_0}=\frac{3m_{3/2}}{2 b s_0}\, .
\end{equation}
The suppression of the gaugino masses with respect to the gravitino mass is usually parametrised by $\varrho,$ which is defined by
\begin{equation}
\varrho:=\frac{16 \pi^2}{m_{3/2}}\frac{F_S}{2 s_0}=\frac{12 \pi^2 }{b}=\frac{3N}{2}\, ,
\end{equation}
where we have used in the last step that the hidden gauge group is $SU(N).$ For $N=4,5$ one then immediately fixes $\varrho=6,7.5.$\\[0.1cm]
\noindent {\bf Scalar Masses:} The moduli mediated contributions to the scalar masses are determined by the coupling of the matter fields $Q_i$ to the hidden sector uplifting-field $X$ in the K\"ahler potential, which can be parametrised as follows
\begin{equation}\label{eq:Kaehler1}
 K_{\rm matter}~=~ Q_\alpha \overline{Q}_\alpha\,  \Bigl[
 1+ \xi_\alpha\, X\, \overline{X} + \mathcal{O} (|X|^4)
\Bigr] \;,
\end{equation}
where $\xi_\alpha$ denotes the effective modular weight of the matter field. As in the type IIB case, the resulting moduli mediated scalar masses are typically not loop-suppressed~\cite{Lebedev:2006qq} and are given by
\begin{equation}
m_{\alpha}^2=m_{3/2}^2 (1-3\xi_\alpha)\, .
\end{equation}
As discussed in~\cite{Krippendorf:2012ir} the modular weights $\xi_\alpha$ need to be distinguished for twisted and un-twisted matter fields.
Typically we find the Higgses, $t_{R}$, $Q_3$ in the untwisted sector and in part of the models also $\tau_R$. So we distinguish two scenarios depending on whether $\tau_R$ is in the twisted or untwisted sector. The modular weight of fields in the untwisted sector is denoted by $\xi_3$ and their moduli mediated mass by $m_3.$ The other matter fields are in the twisted sector with modular weight $\xi_1$ and moduli mediated mass $m_1$.
\begin{center}
\begin{tabular}{c | c }
&  Untwisted sector\\ \hline
Scenario 1 & $H_{u,d},\ Q_3,\ t_R$\\
Scenario 2 & $H_{u,d},\ Q_3,\ t_R,\ \tau_R$
\end{tabular}
\captionof{table}{\footnotesize{The geometric localisation of MSSM fields in the untwisted and twisted sector. The MSSM fields not listed are located in the twisted sector. The modular weight for fields in the twisted sector is $\xi_1$ and for untwisted sector fields $\xi_3.$}\label{tab:scenarios}}
\end{center}

We expect $\xi_3$ to be close to the no-scale value of $1/3$ resulting in almost vanishing soft scalar masses, whereas $\xi_1$ is arbitrary. Note, however, that large negative values of $\xi_1$ would lead to soft-scalar masses for the twisted fields that are not related to the gravitino mass. Although theoretically not excluded, we shall exclude this possibility at this stage as this would require a non-minimal UV construction to achieve such values for $\xi_1$ which is beyond the scope of this article.

Due to the suppression of tree-level moduli mediated contributions, anomaly mediated contributions become important, leading to a mixture of moduli and anomaly mediated soft-masses, referred to as the mirage scheme.  As presented for example in~\cite{Krippendorf:2012ir}, the soft-masses including the anomaly-mediated contributions can be summarised as follows:
\begin{eqnarray}\label{eq:softmassscheme}
 M_a & = & \frac{m_{3/2}}{16 \pi^2}\,\left[\varrho+b_a\,g_a^2\right]\;,\\
 A_{\alpha\beta\delta} & = & 
 \frac{m_{3/2}}{16 \pi^2}\,\left[-\varrho+\left(\gamma_\alpha+\gamma_\beta+\gamma_\delta\right)\right]
 \;,\\
 m_\alpha^2 & = & \frac{m_{3/2}^2}{(16\pi^2)^2}\,\left[\varrho^2\xi_\alpha-\Dot{\gamma}_\alpha
 	+2\varrho\,\left(S+\overline{S}\right)\,\partial_S\gamma_\alpha
	+(1-3\xi_\alpha)(16\pi^2)^2\right]\; ,
\end{eqnarray}
where $\gamma_i$ denote the standard anomalous dimensions as present in anomaly mediated supersymmetry breaking (for a detailed explanation of the notation see~\cite{Falkowski:2005ck} and in particular appendix~A therein). 
\subsection{Summary of UV parameters}\label{sec:summaryuv}
As discussed previously, the value of $\varrho$ is fixed to $\varrho=6,7.5$ depending on whether the hidden sector gauge group is SU(4) or SU(5). The modular weights for the twisted fields $\xi_1$ are a priori undetermined due to the missing UV understanding of their K\"ahler potential. However, the largest mass available for supersymmetry breaking masses is typically the gravitino mass and this equips us in turn with a lower limit for $\xi_1,$ i.e.~$\xi_1>0.$ In contrast, $\xi_3$ for the untwisted fields is expected to be close to the no-scale value $\xi_3=1/3$ as argued in~\cite{Krippendorf:2012ir}. In the following, we discuss different values of $\xi_{1,3}$ in terms of their respective moduli mediated scalar masses $m_{1,3}.$ As the value of $\xi_1$ determines the exact ratio between the gaugino mass and soft scalar masses, the value can become important for reducing the amount of fine-tuning in a given model. Of the remaining free parameters, $B\mu$ can be traded against $\tan\beta$, the ratio of the Higgs vacuum expectation values, while $|\mu|$ can be fixed at the low scale by requiring proper electroweak symmetry breaking. Note that within the MiniLandscape models,  a weak scale $\mu$-term can emerge from an underlying approximate R-symmetry~\cite{Kappl:2008ie}.

\section{Low-energy phenomenology from the heterotic MiniLandscape}
\label{sec:phenoanalysis}
In this section, we study the phenomenological implications of the MiniLandscape models with the soft terms and parameters introduced in the previous section. As already mentioned, we distinguish two phenomenological scenarios, depending on whether $\tau_R$ is located in the twisted (scenario 1) or the untwisted sector (scenario 2). 

\begin{figure}[htp]
\includegraphics[height=8.95cm]{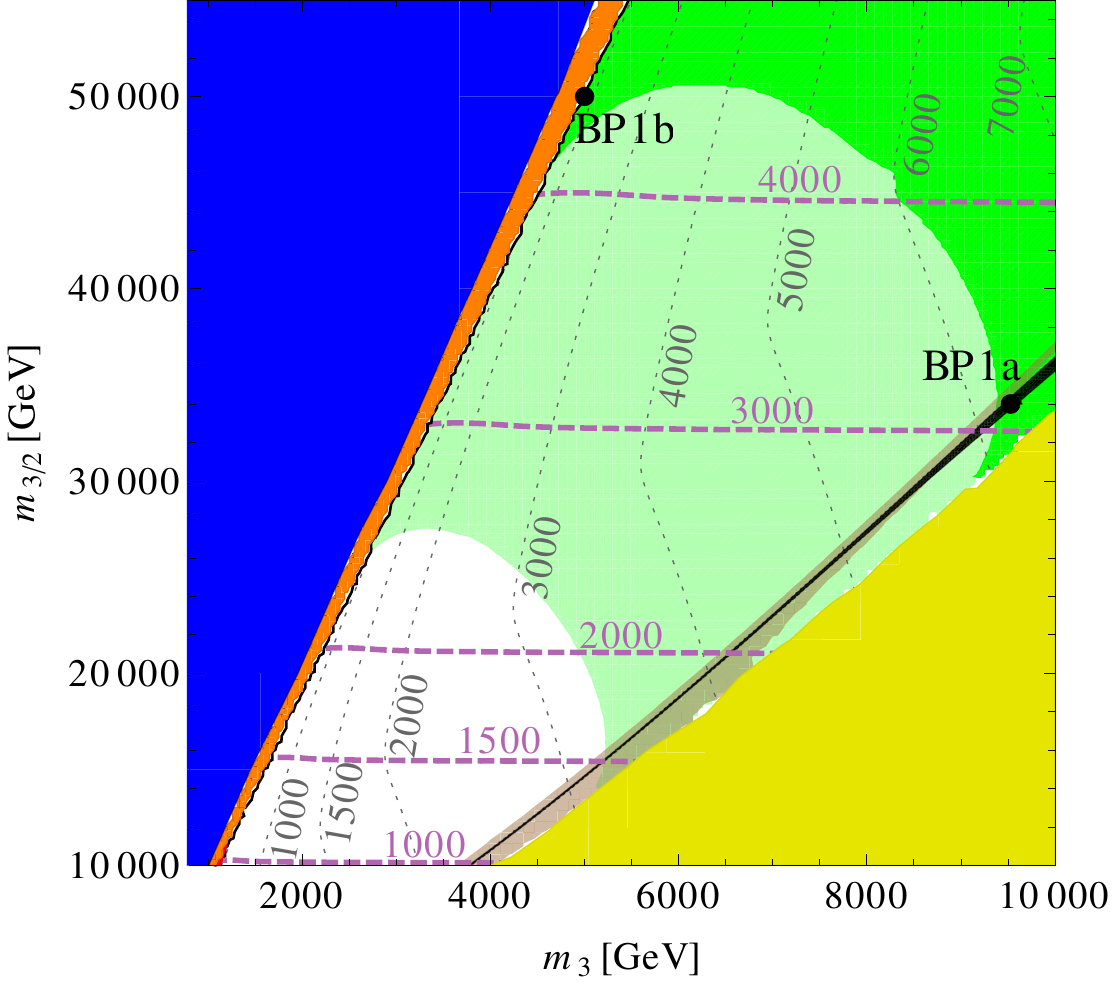} \hspace{0.2cm}\includegraphics[height=7cm]{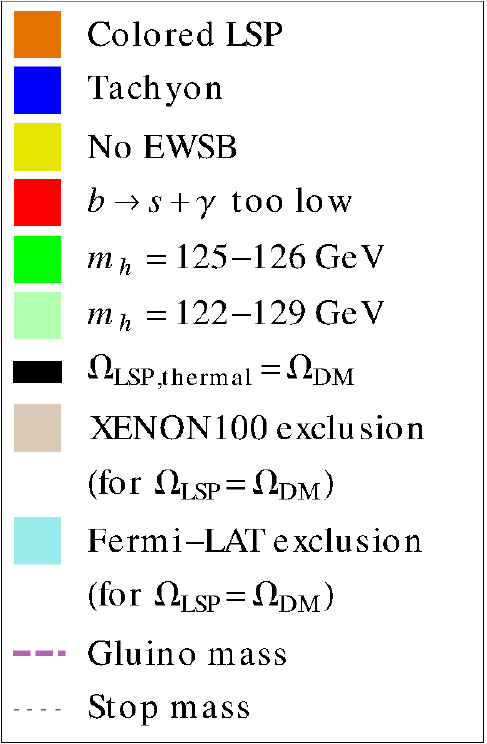}\\[0.5cm]
\includegraphics[height=8.95cm]{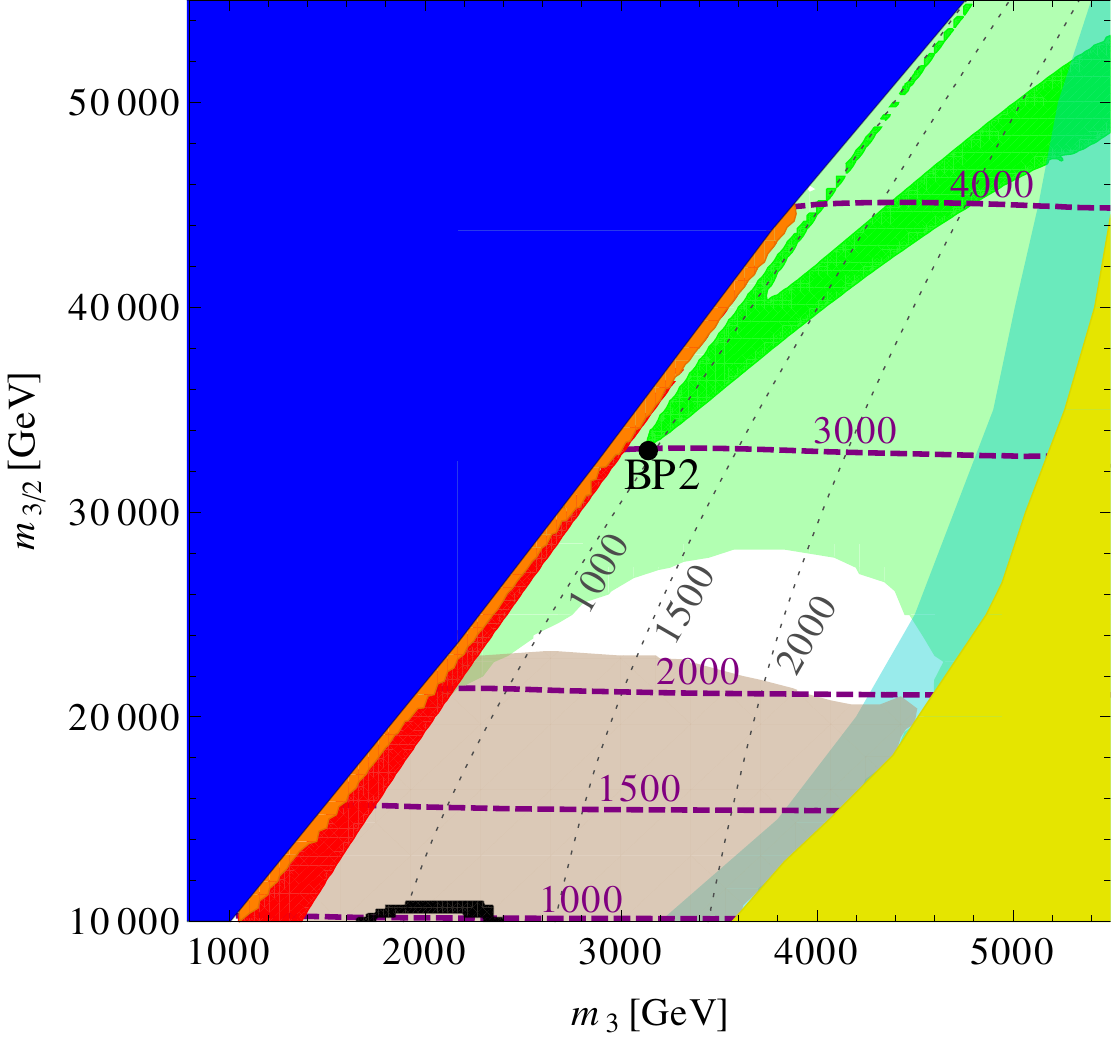}
\captionof{figure}{\footnotesize{Parameter scans for scenario 1 with $\tau_R$ in the twisted sector (upper panel) and scenario 2 with $\tau_R$ in the untwisted sector (lower panel). We have set $m_1=0.5\,m_{3/2}$, $\varrho=7.5$, $\tan\beta=10$ and $\text{sgn}\mu=+$. Contours for gluino and stop mass are depicted together with various phenomenological constraints as described in the box above. In table~\ref{tab:benchmark} we provide the mass spectra and relevant observables for the three benchmark points indicated in the scans.}}
\label{fig:scans}
\end{figure}

For both scenarios we have calculated the low energy mass spectra with the modified version of Softsusy 3.2.4~\cite{Allanach:2001kg} described in~\cite{Badziak:2012rf} which avoids a code problem occurring for light stops. Electroweak precision observables as well as the thermal relic density and the direct detection cross section of the lightest supersymmetric particle (LSP) were determined with MicrOMEGAs~2.4.5~\cite{Belanger:2010gh}. In order to illustrate the main properties of the models, we provide parameter scans in figure~\ref{fig:scans}, where we apply various phenomenological constraints. The resulting superpartner mass spectra and other relevant observables for three benchmark points as indicated in figure~\ref{fig:scans} are shown in table~\ref{tab:benchmark}. Based on the scans we will describe the most important properties of the MiniLandscape models in the remainder of this section.

\begin{table}[h!]
\centering
\begin{footnotesize}
\begin{tabular}{|l|c|c|c|}
\hline
  & \textbf{BP1a} & \textbf{BP1b} & \textbf{BP2} \\
\hline\hline 
\multicolumn{4}{c}{} \\[-4mm]
\multicolumn{4}{c}{Gravitino}\\
\hline
$m_{3/2}$ [TeV] & 34 & 50 & 33\\
\hline\hline
\multicolumn{4}{c}{} \\[-4mm]
\multicolumn{4}{c}{Higgs sector}\\
\hline
$m_h$ [GeV]& 125.1 & 125.4 & 125.0 \\
$m_H$ [TeV]& 9.55 & 6.60 & 2.55 \\ 
$m_a$ [TeV]& 9.55 & 6.60 & 2.55 \\
\hline\hline
\multicolumn{4}{c}{} \\[-4mm]
\multicolumn{4}{c}{Neutralinos}\\
\hline
$m_{\widetilde{\chi}_1}$ [TeV]& 0.941 & 1.586 & 0.493 \\
$m_{\widetilde{\chi}_2}$ [TeV]& 0.959 & 2.17 & 0.503 \\
$m_{\widetilde{\chi}_3}$ [TeV]& 1.08 & 3.78 & 1.02 \\
$m_{\widetilde{\chi}_4}$ [TeV]& 1.51 & 3.78 & 1.43 \\
gaugino fraction $\chi_1$& 7.7\% & 99.97\%  & 0.9\% \\
higgsino fraction $\chi_1$& 92.3\% & 0.03\%  & 99.1\% \\ 
\hline\hline
\multicolumn{4}{c}{} \\[-4mm]
\multicolumn{4}{c}{Charginos}\\
\hline
$m_{\widetilde{\chi}_1^+}$ [TeV]& 0.95 & 2.17 & 0.498 \\
$m_{\widetilde{\chi}_2^+}$ [TeV]& 1.51 & 3.78 & 1.43 \\
\hline\hline
\multicolumn{4}{c}{} \\[-4mm]
\multicolumn{4}{c}{Gluino}\\
\hline
$m_{\widetilde{g}}$ [TeV]& 3.12 & 4.42 & 2.99 \\
\hline\hline
\multicolumn{4}{c}{} \\[-4mm]
\multicolumn{4}{c}{Sfermions 3$^\text{rd}$ generation}\\
\hline
$m_{\widetilde{t}_1}$ [TeV] & 6.22 & 1.591 & 0.95 \\
$m_{\widetilde{t}_2}$ [TeV] & 7.53 & 4.80 & 1.51 \\
$m_{\widetilde{b}_1}$ [TeV] & 7.53 & 1.591 & 1.47 \\
$m_{\widetilde{b}_2}$ [TeV] & 16.9 & 24.7 & 16.5 \\
$m_{\widetilde{\tau}_1}$ [TeV] & 16.6 & 24.3 & 2.90 \\
$m_{\widetilde{\tau}_2}$ [TeV] & 17.1 & 25.2 & 16.5\\
\hline\hline
\multicolumn{4}{c}{} \\[-4mm]
\multicolumn{4}{c}{Sfermions 1$^\text{st}$ and 2$^\text{nd}$ generation}\\
\hline
$m$ [TeV]& $\sim 17$ & $\sim 25$ & $\sim 17$\\
\hline\hline
\multicolumn{4}{c}{} \\[-4mm]
\multicolumn{4}{c}{Dark matter}\\
\hline
$\Omega_{\text{LSP,thermal}} \:h^2$ & 0.1 & 0.1 & 0.03\\
$\sigma_n$ [cm$^2$]& $1.4\cdot 10^{-44}$ & $< 10^{-50}$ & $2.6\cdot 10^{-45}$ \\
\hline\hline
\multicolumn{4}{c}{} \\[-4mm]
\multicolumn{4}{c}{Electroweak observables}\\
\hline
$\text{Br}(b \rightarrow s \gamma)$& $3.3\cdot 10^{-4}$ & $3.2\cdot 10^{-4}$ & $2.9\cdot 10^{-4}$ \\
$\text{Br}(B_s \rightarrow \mu\mu)$ & $3.1\cdot 10^{-9}$ & $3.1\cdot 10^{-9}$ & $3.1\cdot 10^{-9}$\\
\hline\hline
\multicolumn{4}{c}{} \\[-4mm]
\multicolumn{4}{c}{Higgs decays (normalised to SM)}\\
\hline
$h \rightarrow \gamma \gamma$& 1.0 & 1.0  & 1.01 \\
$h \rightarrow gg$& 1.0 & 1.0 & 0.98 \\
\hline
\end{tabular}
\end{footnotesize}
\captionof{table}{\footnotesize{The mass spectrum, electroweak observables, LSP thermal relic density and direct detection cross section for the three benchmark points indicated in figure~\ref{fig:scans}.}}
\label{tab:benchmark}
\end{table}

\subsection{The superpartner spectrum}\label{sec:inverted}

In both scenarios, the sfermions of the twisted sector become very heavy from an LHC perspective. Their natural mass scale is the gravitino mass which typically takes values $m_{3/2}>10\tev$ in the models under consideration (see section~\ref{sec:massbounds}). The sfermions of the untwisted sector are expected to be significantly lighter due to their localization properties. But note that this is also a phenomenological requirement as electroweak symmetry breaking is absent for too large values of $m_3$ (yellow regions in figure~\ref{fig:scans}).

On the other hand, the hierarchy between the untwisted and twisted sector sfermions cannot be too large. This is because the heavy scalars of the twisted sector enter the renormalisation group equations (RGEs) of $m_{\widetilde{Q}_3}$ and $m_{\widetilde{t}_R}$ at the two-loop level. More specifically, heavy twisted sector sfermions tend to reduce the stop masses through the RGE running. If the hierarchy gets too large, the lighter stop becomes the LSP (orange regions in figure~\ref{fig:scans}) or even tachyonic (blue regions).

Since the soft third-generation sfermion masses are non-universal in the heterotic MiniLandscape, the heavy twisted sector sfermions may enter the RGEs of $\widetilde{Q}_3$ and $\widetilde{t}_R$ also at the one-loop level through the combination:
\begin{equation}
 S_Y = m_{H_u}^2-m_{H_d}^2+\sum_{i=1}^3[m_{\widetilde{Q}_i}^2-2m_{\widetilde{U}_i}^2+m_{\widetilde{D}_i}^2-m_{\widetilde{L}_i}^2+m_{\widetilde{E}_i}^2] \,.
\end{equation}
The value of $S_Y$ is the main source of phenomenological differences between the two scenarios. For $\tau_R$ in the untwisted sector (scenario 2) $S_Y$ vanishes and does not affect the low-energy spectrum. As a result, for heavy enough twisted sector sfermions both stops can be relatively light and strongly mixed (see section~\ref{sec:Higgs} and~\cite{Badziak:2012rf}). On the other hand, for $\tau_R$ in the twisted sector (scenario 1) $S_Y=m_1^2-m_3^2$ at the high scale; so it is typically positive and very large since generically $m_1\sim{\mathcal{O}}(m_{3/2})\gg m_3$. The contribution from $S_Y$ to the electroweak scale soft scalar masses is determined by the hypercharge assignment and is approximately given by
\begin{equation}
 m^2_i = -0.05 Y_i S_Y \,,
\end{equation}
where $Y_i$ is the hypercharge of the sfermion $i$. In particular, $S_Y$ gives a positive contribution to $m_{\widetilde{U}}^2\simeq0.035 S_Y$ and compensates the negative two-loop effect, $(m_{\widetilde{U}}^2)^{\text{2-loop}}\simeq-0.02 m_1^2$. The contribution to $m_{\widetilde{Q}}^2\simeq-0.008 S_Y$ is negative and relatively small. In consequence, in scenario 1 only one stop can be light and the left-right stop mixing is smaller than in scenario 2. This can be seen in the benchmark point BP1b in table~\ref{tab:benchmark}  where $\widetilde{t}_1$ becomes relatively light through the RGE effects, while $\widetilde{t}_2$ stays heavy. Moreover, since $m_{\widetilde{Q}}$ is significantly smaller than $m_{\widetilde{U}}$ at the electroweak scale, the lighter stop is mostly left-handed, in contrast to conventional models such as the CMSSM, in which $m_{\widetilde{U}}< m_{\widetilde{Q}}$ typically holds. For scenario 2, both stops may become light as in the benchmark scenario BP2.

Another consequence of a large and positive $S_Y$ in scenario 1 typically is a heavy higgsino which follows from the fact that $S_Y$ gives a large negative contribution to the electroweak scale value of $m_{H_u}^2\simeq-0.025 S_Y$ and $\mu^2\simeq -m_{H_u}^2$ from the condition of proper electroweak symmetry breaking. Therefore the LSP is typically a bino-like neutralino in scenario 1 (cf. benchmark point BP1b). However, there is some parameter space close to the ``No EWSB'' region where the higgsino gets light due to accidental cancellations in the RGE of $m_{H_u},$ which can be seen in the benchmark point BP1a. In scenario 2, the $\mu$ parameter is suppressed due to the vanishing $S_Y$ and due to generic cancellations in the RGE of $m_{H_u}$ which we shall discuss in more detail in the context of reduced fine-tuning in section~\ref{sec:redfine-tuning}. As a result, the LSP is typically an almost pure higgsino in scenario 2 (cf. benchmark point BP2).

\subsection{The Higgs sector}\label{sec:Higgs}

In MSSM models, a Higgs mass of $m_h\simeq 126\gev$ as measured by ATLAS and CMS~\cite{atlashiggs,cmshiggs}, can only be accommodated in the presence of large loop corrections to $m_h$. Applying the decoupling limit on the MSSM Higgs bosons and assuming $\tan\beta \gg 1$, one finds~\cite{Haber:1996fp}
\begin{equation}
 m_h\simeq M_Z^2 + \frac{3\,g_2^2\,m_t^4}{8\,\pi^2\,m_W^2}\left[ \log\left(\frac{m_{\widetilde{t}}^2}{m_t^2}\right)+\frac{A_t^2}{m_{\widetilde{t}}^2}\left(1-\frac{A_t^2}{12\,m_{\widetilde{t}}^2}\right)\right]\;,
\end{equation}
where we included the dominant one-loop contributions from the top/stop sector and introduced $m_{\widetilde{t}}=\sqrt{m_{\widetilde{t}_1} m_{\widetilde{t}_2}}$\,. A sufficiently large $m_h$ requires either heavy stops ($m_{\widetilde{t}}\gg 1\tev$) or sizeable stop mixing. For a given stop mass, $m_h$ is maximized for $|A_t|=\sqrt{6}\,m_{\widetilde{t}}$.

\begin{figure}[h!]
\begin{center}
\includegraphics[height=4.8cm]{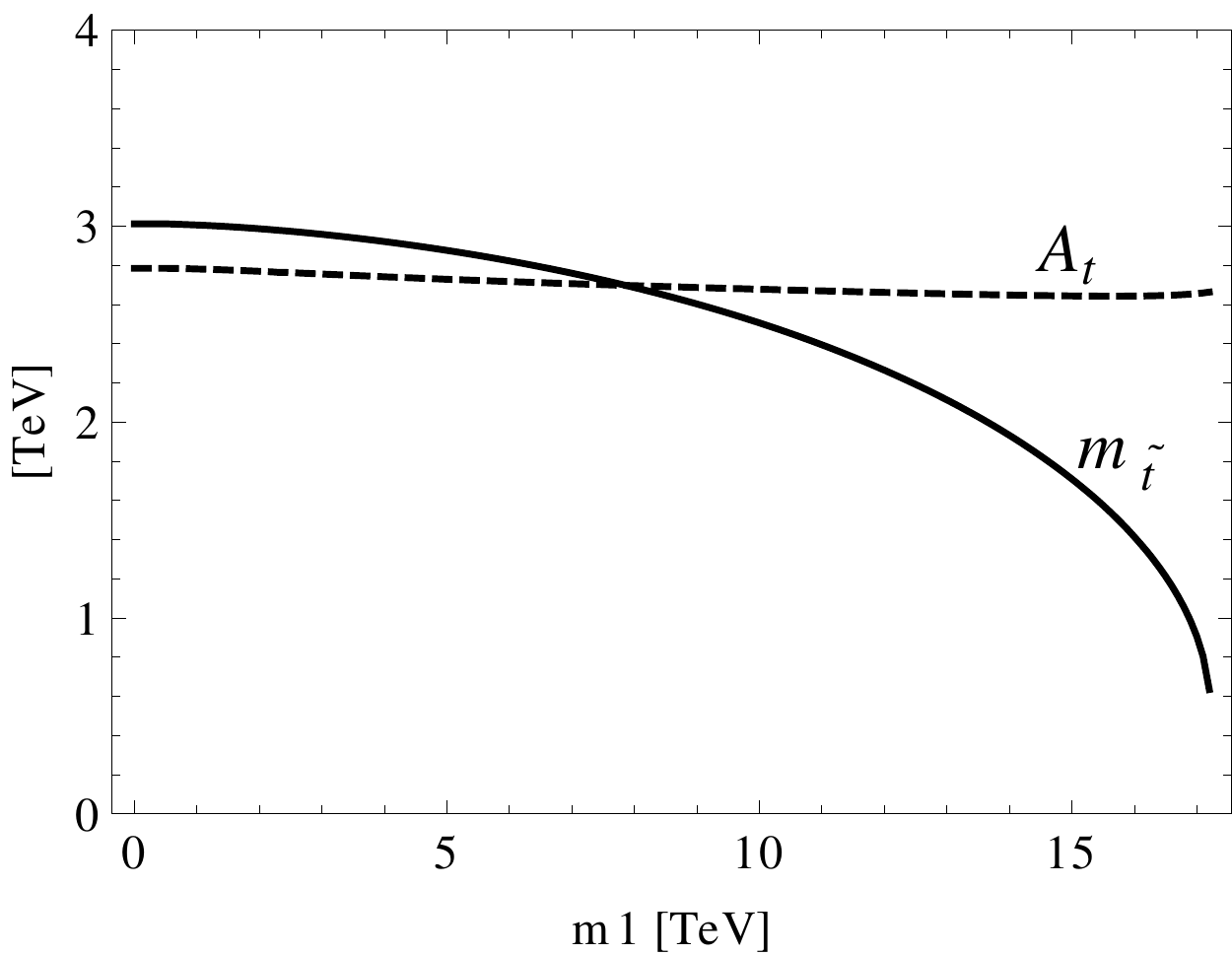}\hspace{10mm}
\includegraphics[height=4.8cm]{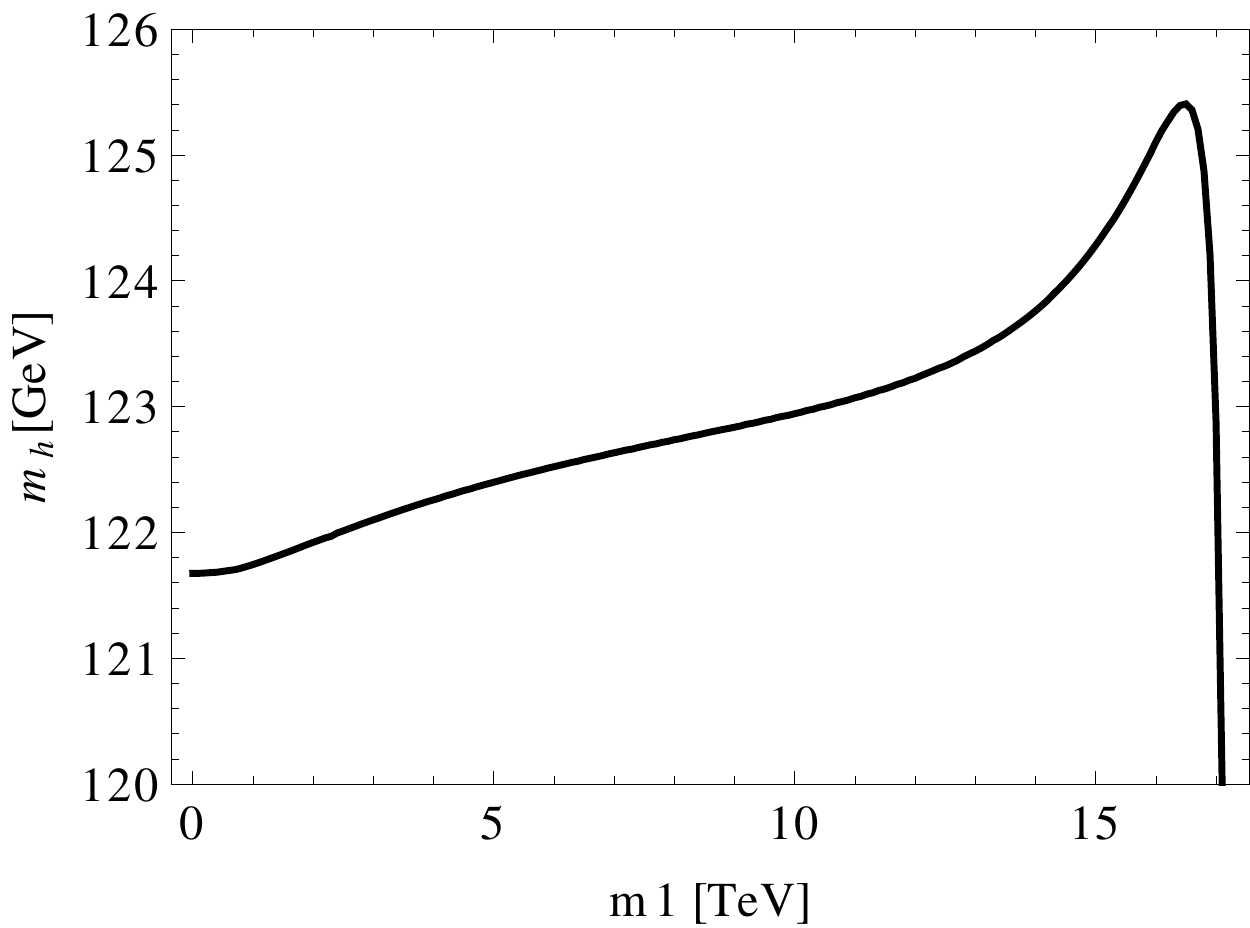} 
\end{center}
\captionof{figure}{\footnotesize{Stop mass $m_{\widetilde{t}}=\sqrt{m_{\widetilde{t}_1} m_{\widetilde{t}_2}}$  and top trilinear coupling $A_t$ (left panel) as well as the Higgs mass $m_h$ (right panel) as a function of the twisted sector sfermion mass $m_1$. We have fixed $m_{3/2}=35\tev$, $m_3=3\tev$, $\varrho=7.5$, $\tan\beta=10$, $\text{sgn}\mu=+$ and assumed that $\tau_R$ resides in the untwisted sector (scenario 2).}}
\label{fig:stopmixing}
\end{figure}

In the MiniLandscape models, there exists the interesting possibility to generate large stop mixing through RGE effects (see also~\cite{Badziak:2012rf} for more details on stop-mixing). As discussed previously, the RGE running of stops is affected by the heavy twisted sector sfermions. With increasing hierarchy in the scalar sector, $m_{\widetilde{t}}$ is reduced, while the trilinear coupling $A_t$ is insensitive to the choice of $m_1$ (see left panel of figure~\ref{fig:stopmixing}). Therefore, the stop mixing grows with increasing $m_1$. The corresponding effect on the Higgs mass is shown on the right panel of figure~\ref{fig:stopmixing}. The Higgs mass gets larger until a maximum is reached at $|A_t|\simeq\sqrt{6}\,m_{\widetilde{t}}$. Beyond this so-called ``maximum-mixing'' case, the Higgs mass decreases again.

In the scans in figure~\ref{fig:scans}, we have marked the parameter space consistent with the observed Higgs mass ($m_h=125-126\gev$) in green. The light green region becomes viable if we include an additional theoretical uncertainty of $3\gev$ on the calculation of the Higgs mass. The RGE effects on the stop masses, which we just discussed, show up for light stop masses. This can be seen by the fact that the Higgs mass grows towards the region where the stop becomes the LSP. The effects are more pronounced in scenario~2 where $\tau_R$ resides in the untwisted sector. The reason is, that in scenario 1 only the left-handed stop gets light through the RGE effects due to the non-zero $S_Y$-parameter, i.e. maximal stop mixing cannot be reached (see section~\ref{sec:inverted}). In the regime with relatively large $m_3$, which is only accessible in scenario 1, the effects of stop mixing become negligible as $m_{\widetilde{t}}\gg|A_t|$ and the Higgs mass simply increases with growing $m_{\widetilde{t}}$.

Turning to the decay properties of the light Higgs $h$ in models of the MiniLandscape, we find them to be very similar to the Standard Model. This is because the decoupling limit on the Higgs sector generically applies where the couplings of $h$ are Standard Model-like. However, in the presence of light stops, the radiative decays $h\rightarrow \gamma\gamma,gg\,$ may get affected by the interference of the stop loop with the relevant Standard Model processes~\cite{Djouadi:1998az}. For large left-right stop-mixing, the stop loop enters the amplitude with the same sign as the $W$ loop (which dominates the decay to photons), but with opposite sign as the top loop (which dominates the decay to gluons). Therefore, light stops tend to enhance $\text{Br}(h\rightarrow \gamma\gamma)$, but reduce $\text{Br}(h\rightarrow gg).$ The latter also suppresses the Higgs production through gluon fusion. The Higgs production is more affected by the stop loops than $\text{Br}(h\rightarrow \gamma\gamma)$ so the $\gamma\gamma$ production rate, $\sigma(pp\rightarrow h)\times\text{Br}(h\rightarrow \gamma\gamma)$, is reduced as compared to the SM prediction.

We have determined the branching fractions $\text{Br}(h\rightarrow \gamma\gamma,gg)$ including the SUSY contributions from the stop sector as given e.g.\ in~\cite{Djouadi:2005gj}. We find them to be very close to the corresponding SM branching fractions in the parameter regions which satisfy the phenomenological constraints and are consistent with the Higgs mass bounds. In the benchmark scenarios of table~\ref{tab:benchmark}, the deviation is at the percent level for BP2, while it is totally negligible for BP1a and BP1b. Only in the region which is already excluded either by a stop LSP or a too light $m_h,$ the branching fractions $\text{Br}(h\rightarrow \gamma\gamma,gg)$ may deviate by more than ten percent from the Standard Model values.

\subsection{Mirage pattern in the gaugino sector}

The gaugino pattern in the MiniLandscape models is fully predictable and markedly different from standard schemes like the CMSSM. As discussed in section~\ref{sec:scalesinminilandscape}, the gaugino masses $M_a$ receive comparable contributions from modulus and anomaly mediation, the latter being non-universal among the $M_a$. As the splitting of the gaugino masses at the high scale is fixed by the same $\beta$-function coefficients which determine their RGE running, the gaugino masses unify at an intermediate so-called ``mirage scale''. At the low scale, the gaugino hierarchy in the MiniLandscape models is the same as in the CMSSM (bino lightest, gluino heaviest), but their spectrum is considerably compressed. Up to two-loop corrections, the pattern is completely fixed by the parameter $\varrho$ which determines the relative size of modulus and anomaly mediated contributions to the gaugino masses. As discussed in section~\ref{sec:topdownnatural}, $\varrho$ can take the discrete values $\varrho=6$ and $\varrho=7.5$ depending on whether the hidden sector gauge group is a SU(4) or a SU(5). Therefore, we obtain a prediction for the physical gaugino mass pattern in the MiniLandscape models which is compared to the CMSSM in table~\ref{tab:gauginopattern}. Note that the compressed gaugino spectrum has important implications for SUSY searches at the LHC as discussed e.g.\ in~\cite{Dreiner:2012gx,Dreiner:2012sh}.
\begin{table}[h!]
\centering
 \begin{tabular}[h]{|cl|c|}
\hline
\multicolumn{2}{|c|}{\textbf{Model}} & $\boldsymbol{m_{\widetilde{B}}:m_{\widetilde{W}}:m_{\widetilde{g}}}$\\
\hline\hline 
&& \\[-4mm]
MiniLandscape&($\varrho=6$) & $1\,:\,1.3\,:\,2.6$ \\
&($\varrho=7.5$) & $1\,:\,1.4\,:\,2.9$\\
\hline\hline
& & \\[-4mm]
\multicolumn{2}{|c|}{CMSSM} & $1\,:\,1.9\,:\,5.3$\\ \hline
\end{tabular}
\captionof{table}{\footnotesize{Gaugino mass pattern in mirage mediation for the two realistic choices of $\varrho.$ For comparison the pattern in the CMSSM is also shown.}}
\label{tab:gauginopattern}
\end{table}

\subsection{Lower limit on the gravitino and gluino mass}\label{sec:massbounds}
As can be seen for example in figure~\ref{fig:scans}, the measured Higgs mass sets a lower bound on the gravitino mass. This can easily be understood as the gravitino sets the overall scale of the soft terms which enter the loop contribution to $m_h$. In order to determine the limit on the gravitino mass we have chosen the remaining free parameters ($m_3$, $m_1$, $\tan\beta$, $\varrho$) within the theoretical limits as described in section~\ref{sec:summaryuv} such that for a given Higgs mass the gravitino mass is minimised. 

For example, an increase in the twisted sector scalar masses $m_1$ compared to the gravitino mass generally speaking leads to lower $m_{3/2}$ consistent with the observed Higgs mass. If we now require $m_h>125\gev$, we obtain the limit 
\begin{equation}
m_{3/2} > 15\tev
\end{equation}
 which is saturated for scenario~1 with $\tau_R$ in the twisted sector (in scenario 2 the constraint is even stronger). Note that such large gravitino masses are desirable from a cosmological perspective as the gravitino would decay before primordial nucleosynthesis, thus considerably ameliorating the gravitino problem~\cite{Weinberg:1982zq}.

This bound on the gravitino mass can be directly translated into a lower limit on the gluino mass which reads
\begin{equation}
\label{gluinobound}
 m_{\widetilde{g}} > 1.2\tev\;.
\end{equation}
The above bound incidentally coincides with the recent lower bound on the gluino mass from the ATLAS search for gluino pair production in final states with multiple b-jets~\cite{ATLAS-CONF-2012-145} which was found at $ m_{\widetilde{g}} \simeq 1.2\tev$ for our ratio of gluino to LSP mass $m_{\widetilde{\chi}_1}\lesssim 0.4 \,m_{\widetilde{g}}$ (cf. table~\ref{tab:gauginopattern}). Even though this limit was set under the assumption of a simple gluino decay chain, $\widetilde{g}\rightarrow t\bar{t}\widetilde{\chi}_1$ or $\widetilde{g}\rightarrow b\bar{b}\widetilde{\chi}_1$, we expect a rather similar bound\footnote{The exact limit may be slightly weaker than in the simplified model because the gluino usually does not decay directly to the LSP.} in the heterotic MiniLandscape scenario since typically the gluino decays via off-shell stops to $t\bar{t}$ pairs associated by (not necessarily the lightest) neutralino or to $b\bar{t}$ (or $\bar{b}t$) and a chargino so that the final states are rich in b-jets. 
Although the lower bound~\eqref{gluinobound} is very close to the current experimental lower gluino mass limit, large fractions of the parameter space favour a gluino mass in the multi TeV range, i.e.~in particular above the energy accessible at the LHC (see e.g.~\cite{Baer:2012vr} for the gluino mass reach of the LHC). Note, however, that the constraints on $m_{3/2}$ and $m_{\widetilde{g}}$ get weaker if we take into account the theoretical uncertainty of the Higgs mass calculation by the spectrum calculator. For instance, if we allow for a Higgs mass $m_h=122\gev$ (corresponding to a theoretical uncertainty of 3 GeV), we find the limit $m_{3/2}>7.5\tev$ which implies $m_{\widetilde{g}}>600\gev$.
However this would be below the current LHC search limits for gluinos. 

\subsection{Flavour constraints}

Flavour constraints on the MiniLandscape models mainly arise from the decay $b\rightarrow s \, \gamma$. The present experimental value of the branching fraction~\cite{Asner:2010qj}
\begin{equation}
\text{Br}(b\rightarrow s \, \gamma)=\left(3.55 \pm 0.24 \pm 0.09\right)\cdot 10^{-4}
\end{equation}
is consistent with the Standard Model expectation $\text{Br}^\text{SM}(b\rightarrow s \, \gamma)=(3.2\pm 0.2)\cdot 10^{-4}$~\cite{Misiak:2006zs}. Large SUSY contributions to $b\rightarrow s \, \gamma$ may be generated through charged Higgs or chargino/squark loops. In the MiniLandscape models, mainly the chargino/stop contribution is relevant as these fields are relatively light in some part of the parameter space (while the charged Higgs bosons are typically heavy). This is very similar as in the inverted hierarchy models discussed in~\cite{Badziak:2012rf}. We have used MicrOMEGAs 2.4.5~\cite{Belanger:2010gh} to calculate $\text{Br}(b\rightarrow s \, \gamma)$ for the MiniLandscape models, the corresponding exclusions on the parameter space are shown in figure~\ref{fig:scans}. As it is difficult to estimate the theoretical uncertainty of the calculation by MicrOMEGAs, we have chosen a rather conservative interval of $3\sigma$ around the experimental central value to obtain the bound. As can be seen in figure~\ref{fig:scans}, a small region of parameter space which otherwise would be viable is indeed excluded by the $b\rightarrow s \, \gamma$ constraint. The higgsino-like chargino is typically lighter in scenario 2 ($\tau_R$ in the untwisted sector) and, correspondingly, the exclusion is stronger for this case.
Note that the SUSY contribution to $b\rightarrow s \, \gamma$ grows with increasing $\tan\beta$, and that it would flip sign for a negative choice of $\mu$.

We have verified that within the MiniLandscape models, the SUSY effects on $B_s\rightarrow \mu\mu$ are negligible unless for very large $\tan\beta$. In addition, the anomalous magnetic moment of the muon is not affected due to the sleptons being heavy.

\subsection{Reduced fine-tuning}
\label{sec:redfine-tuning}

The absence of SUSY signals at the LHC as well as the rather large mass of the Higgs boson $m_h\simeq 126\gev$ seem to prefer the superpartner mass scale to be considerably above the weak scale. On the other hand, heavy superpartners threaten the naturalness of supersymmetric theories as the mass of the $Z$ boson is connected to the scale of the soft terms. Any splitting between the electroweak scale and the scale of the soft terms requires fine-tuning; this is the so-called little hierarchy problem of the MSSM. 

The little hierarchy problem can considerably be ameliorated within the models from the heterotic MiniLandscape, which we would like to specify in the following. As discussed in section~\ref{sec:Higgs}, the hierarchy in the scalar sector may induce large stop mixing through the RGE running. The effect is especially strong in scenario 2 ($\tau_R$ in the untwisted sector), where a Higgs mass $m_h\sim 126\gev$ can still be realised with both stops at around $1\tev$. In this scenario there exists a mechanism to reduce the fine-tuning even further. To illustrate this, it is instructive to express $M_Z^2$ in terms of the high scale parameters. For scenario 2, we find (for $\tan{\beta}=10$)
\begin{equation}\label{eq:zmass}
M_Z^2\simeq 2.8\, M_3^2 - 0.4\, M_2^2 - 0.7\,A_t\,M_3 + 0.2\, A_t^2 - 0.1\,m_3^2 - 0.01\,m_1^2 - 2 \mu^2\;.
\end{equation}
The fine-tuning can then be defined as the sensitivity of $M_Z$ with respect to a certain high scale parameter.

One might think that the fine-tuning is always dominated by the gluino due to the large coefficient in front of $M_3$. Note, however, that the twisted sector sfermions are much heavier than the other superpartners in the considered scheme. Their contribution to $M_Z$ is therefore non-negligible despite the small coefficient in front of $m_1$. In addition, $M_3$ and $m_1$ are related as they both depend on the gravitino mass (cf.~equation~\eqref{eq:softmassscheme})
\begin{equation}
 M_3 = \frac{m_{3/2}}{16\,\pi^2}\,(\varrho-1.5)\;, \qquad m_1 = \sqrt{1-3\,\xi_1}\:m_{3/2}\;.
\end{equation}
As the parameter $\varrho$ may only take the discrete values $6$ or $7.5$ (depending on the hidden sector gauge group), it can easily be verified that, especially for $\varrho=7.5,$ there arise strong cancellations between the gluino and sfermion contributions to $M_Z.$ In this case the fine-tuning is considerably reduced. The most favourable situation is achieved for a small, but non-zero $\xi_1$ which constitutes the best compromise between a low fine-tuning with respect to $m_{3/2}$ and a low fine-tuning with respect to $\xi_1$. Note that such cancellations do not arise if $\tau_R$ resides in the twisted sector (scenario 1) as in this case $M_Z$ receives an additional contribution $\sim 0.05 m_1^2$ due to the non-zero $S_Y$ parameter (see section~\ref{sec:inverted}). Therefore, in the MiniLandscape models, the fine-tuning is generically lower if $\tau_R$ is in the untwisted sector.

\subsection{Dark matter}
\paragraph*{Thermal production}

In scenario 1 with $\tau_R$ in the twisted sector, the higgsino is typically quite heavy and, consequently, we obtain a mostly bino LSP in wide regions of parameter space. As there exists no bino-bino-gauge boson vertex at tree-level, its annihilation cross section is suppressed. Therefore, we encounter thermal overproduction of dark matter in most of the parameter space shown in the upper panel of figure~\ref{fig:scans}. However there are two exceptions:
\begin{itemize}
\item If the stop becomes light through RGE running, stop coannihilations may suppress the bino abundance. Indeed, there is a very thin stripe at the border of the stop LSP region where the thermal LSP density is equal or less than the dark matter density. The benchmark point BP1b in table~\ref{tab:benchmark} is chosen from this region.
\item On the other hand, the individual contributions in the RGE of $m_{H_u}$ may cancel accidentally in which case the higgsino gets lighter. In the upper panel of figure~\ref{fig:scans}, a light higgsino occurs close to the region where electroweak symmetry breaking is absent. For a sufficient higgsino fraction, the LSP abundance again (under)matches the dark matter abundance (cf. benchmark point BP1a).
\end{itemize}
In scenario 2 with $\tau_R$ in the untwisted sector, the LSP is typically a higgsino with a small bino and wino admixture. In the early universe the higgsinos undergo very efficient annihilations into gauge bosons and third generation quarks. Their effective cross section is further enhanced by coannihilation processes including the charged higgsinos. Therefore, the thermal higgsino density $\Omega_{\text{LSP,thermal}}$ is generically suppressed.
In the lower panel of figure~\ref{fig:scans} we thus find $\Omega_\text{LSP,thermal}<\Omega_\text{DM}$, where $\Omega_\text{DM}\,h^2\simeq 0.1$ is the dark matter density (cf. benchmark point BP2). This holds except for a tiny region at low $m_{3/2}$ where the LSP contains a considerable bino fraction.

\paragraph*{Non-thermal production and dilution}

As in any locally supersymmetric theory, the energy content of the universe may be affected by the presence of moduli fields. We assume that the matter field which dominantly breaks supersymmetry decouples from the low-energy theory (in the toy model presented in appendix~\ref{sec:detrho}, $X$ receives a large mass through the effective $(X\bar{X})^2$ term in the K\"{a}hler potential, see also~\cite{Dine:1983ys,Coughlan:1984yk,Greene:2002ku}). Therefore, we merely have to deal with the dynamics of the dilaton. In the early universe, the latter gets displaced from its zero temperature minimum by inflation~\cite{Coughlan:1983ci} and by finite temperature effects during the reheating phase~\cite{Buchmuller:2004xr}. The dilaton amplitude at reheating $\delta s_\text{RH}$ due to thermal effects can be approximated as
\begin{equation}\label{eq:thermaldisplacement}
 \delta s_\text{RH}\sim \frac{T_R^4}{m_s^2\,M_P}\,,
\end{equation}
where $T_R$ denotes the reheating temperature, $m_s$ the dilaton mass and $M_P$ the Planck mass. The amplitude induced by inflation depends on the details of the inflationary model.\footnote{Note that, assuming the mechanism of dilaton stabilization we employ here, strong constraints on the model of inflation arise from the requirement that the dilaton does not get destabilised during the inflationary phase (see e.g.~\cite{Kallosh:2004yh}).} Subsequent to reheating, the dilaton undergoes coherent oscillations. The corresponding energy density decreases as $a^{-3}$ where $a$ is the scale factor of the universe. Especially, it redshifts slower than radiation and may contribute significantly to, or even dominate the energy content of the universe at late times. By its decay, the dilaton produces Standard Model fields and superpartners, the latter cascading to neutralino LSPs. Therefore, we are left with a non-thermal contribution to the dark matter density. Depending on the dilaton mass and energy density (prior to decay), the so-produced neutralino LSPs may considerably reduce their abundance through annihilations if the corresponding cross section is large enough.

If we assume that the dilaton density never dominates the energy content of the universe, and that neutralino annihilation after dilaton decay is negligible, the non-thermal neutralino relic density can be estimated as
\begin{equation}
 \Omega_\text{LSP,non-thermal} \sim \frac{m_\chi\,\mathcal{S}_0}{3\,H_0^2\,M_P^2}\,\frac{m_s\,{(\delta s_\text{RH})}^2}{T_R^3}\,,
\end{equation}
where $m_\chi$ denotes the mass of the neutralino LSP, $\mathcal{S}_0\simeq 2900\cm^{-3}$ the present entropy density and $H_0=71\km\s^{-1}/\text{Mpc}$ the Hubble parameter.
The total dark matter density then simply reads $\Omega_\text{LSP}=\Omega_\text{LSP,thermal}+\Omega_\text{LSP,non-thermal}$. Given a dilaton mass of $\mathcal{O}(1000\tev)$ and considering only the thermal effects (cf.~equation~\eqref{eq:thermaldisplacement}) $\Omega_\text{LSP,non-thermal}$ is in the range of the dark matter density for $T_R\sim10^8-10^9\gev$. 

If the dilaton dominates the energy content at its decay, it dilutes the thermal dark matter abundance. However, this scenario typically suffers from the overproduction of non-thermal dark matter and/or from a moduli induced gravitino problem~\cite{Endo:2006zj}.

The settings with $\tau_R$ in the untwisted sector typically come with a higgsino LSP, i.e. $\Omega_\text{LSP,thermal}<\Omega_\text{DM}$. Non-thermal higgsinos may easily account for the remaining fraction of the dark matter. The class of models with $\tau_R$ in the twisted sector, which typically have a bino LSP, suffer from thermal overproduction of dark matter unless in the small parameter regions with stop coannihilations or bino/higgsino mixing. In the remaining parameter space, a consistent picture with bino dark matter arises only if the thermal abundance is sufficiently diluted by modulus decay. In turn, this causes problems associated with the regeneration of binos and gravitinos (see above). Nevertheless, we should not completely exclude this possibility as these follow-up problems may be solved through minimal extensions of the hidden sector~\cite{Dine:2006ii}.

\paragraph*{Direct detection}

In the following, we will consider the constraints which arise from direct dark matter detection. In the considered models, the scattering of the lightest neutralino off nucleons is typically dominated by the exchange of the light Higgs. The corresponding cross section can be written as\footnote{We neglect the small differences between the masses of the neutron and proton.}
\begin{equation}
 \sigma_n \simeq \frac{4 \, m_n^4}{\pi}  \,f_q^2\,\left(f^n_u + f^n_d + f^n_s + \frac{6}{27} \,f^n_G \right)^2\;.
\end{equation}
Here $f^n_u$, $f^n_d$, $f^n_s$ and $f^n_G$ specify the up-, down-, strange-quark and gluon contribution to the nucleon mass $m_n$.\footnote{Note that the quantity $f^n_s$ is subject to large experimental uncertainties. In order not to overestimate the direct detection constraints, we made the rather conservative choice $f^n_s=0.13$ consistent with~\cite{Gasser:1990ce}. Larger values of $f^n_s$ were suggested by~\cite{Pavan:2001wz}.} The effective neutralino quark coupling divided by the quark mass reads
\begin{equation}
 f_q = \frac{g_{h\chi_1\chi_1}}{\sqrt{2} \, v_\mathrm{EW}}\,\frac{1}{m_h^2}\,,
\end{equation}
with $v_\mathrm{EW}$ being the Higgs vacuum expectation value and the neutralino-Higgs coupling $g_{h\chi_1\chi_1}$ can be taken from~\cite{Rosiek:1995kg} for instance. Important for us is that $g_{h\chi_1\chi_1}$ vanishes in the limit of a pure gaugino or higgsino LSP, while it becomes large for a strongly mixed state. Therefore, we expect strong direct detection signals, whenever the mass splitting between gauginos and higgsinos is small.

We have systematically calculated $\sigma_n$ with MicrOMEGAs which automatically takes into account further sub-dominant contributions to $\sigma_n$ arising e.g. from squark exchange in the s-channel. On these results for the cross sections, we applied the constraints from the XENON100 direct dark matter search~\cite{Aprile:2012nq}. The corresponding exclusions on the parameter space of the considered models are indicated in figure~\ref{fig:scans}. Note that in order to apply the constraints, we have assumed that the neutralino LSPs make up the total dark matter density ($\Omega_\text{LSP}=\Omega_\text{DM}$), irrespective whether they have to be produced thermally or non-thermally. In case the neutralinos only account for a sub-dominant fraction of the dark matter, the constraints would become correspondingly weaker.

In scenario 2 with $\tau_R$ in the untwisted sector, $\sigma_n$ is generically sizeable. The LSP is dominantly higgsino, but through the non-negligible bino and wino admixture the cross section with nucleons is enhanced as described above. The region with $m_{3/2}\leq 20\tev$ where the gauginos are rather light, is already excluded by the latest XENON100 search. The entire parameter space shown in the lower panel of figure~\ref{fig:scans} is in reach of the next generation direct detection experiments like XENON1T.

If $\tau_R$ resides in the twisted sector (scenario 1), the cross section $\sigma_n$ is typically suppressed. Only close to the region were electroweak symmetry breaking is absent, the higgsino becomes sufficiently light to induce a sizeable gaugino-higgsino mixing, increasing $\sigma_n$. In the upper panel of figure~\ref{fig:scans}, a large part of the region where the thermal neutralino abundance matches the dark matter abundance is therefore excluded by XENON100. In the remaining parameter space, the higgsino becomes rather heavy and decouples. Especially, in the region with bino stop coannihilations, the cross section $\sigma_n$ is so small that it is not even in reach for the next generation direct detection experiments.

\paragraph*{Indirect detection}

Dark matter annihilations in the galactic halo or within dense substructures give rise to cosmic rays which can potentially be observed in the vicinity of the earth. As the measured fluxes of antiprotons and gamma rays are consistent with astrophysical backgrounds, limits on the dark matter annihilation cross section can be set (see e.g.~\cite{Donato:2008jk,Kappl:2011jw,Ackermann:2011wa}). Most relevant for the MiniLandscape models are the constraints from the search for gamma rays from dwarf spheroidal galaxies performed by the Fermi-LAT collaboration~\cite{Ackermann:2011wa}. These are applied in figure~\ref{fig:scans} and determine the Fermi-LAT excluded region (cyan). Scenarios with light higgsinos are excluded as they yield strong annihilations into $W$ and $Z$ boson pairs which induce a too large flux of photons. However, only the models with $\tau_R$ in the untwisted sector (scenario 2) have a light higgsino and are constrained by Fermi.\footnote{In scenario 1 there is an extremely thin band of parameter space with a light higgsino which is excluded by Fermi at the border of the ``No EWSB'' region (yellow). We do not show this in figure~\ref{fig:scans} to avoid clutter.} Note also that for the exclusion to hold, we have assumed that the entire dark matter is in the form of the lightest neutralino which requires non-thermal production.

\section{Conclusions}
\label{sec:conclusions}
%

The MiniLandcape models of heterotic orbifold compactifications give rise to a very predictive MSSM soft mass pattern. In particular, the interplay between supersymmetry breaking and dilaton stabilisation fixes the ratio of the gaugino masses and their relation to the gravitino mass. In the scalar sector, the distinction between bulk fields (untwisted sector) and localised fields (twisted sector) induces an inverted hierarchy, a scheme also known as {\it natural SUSY.} The sfermions of the first two generations as well as $\widetilde{L}_3$ and $\widetilde{b}_R$ receive masses in the multi-TeV range, all of which are well beyond the LHC reach. The stops and Higgs fields as well as -- depending on the construction -- $\widetilde{\tau}_R$ remain lighter.

Experimental constraints considerably restrict the heterotic MSSM models. Especially we find that, in order to accommodate the observed Higgs mass, a gluino with $m_{\widetilde{g}}>1.2\tev$ is required. In most of the viable parameter space the gluino is substantially heavier than the lower bound, implying that the gluino may not be accessible at the LHC. Despite the heavy gluino, the electroweak fine-tuning can be ameliorated as there tend to be cancellations in the RGE of $m_{H_u}$ between the contributions of the gluino and the sfermions of the two heavy families.

In some part of the parameter space the stops are as light as $\sim 1\tev$, where a sufficiently large Higgs mass follows from strong stop mixing. This mixing is generated by a two-loop RGE effect which is driven by the heavy twisted sector sfermions and enhances $A_t$ relative to the stop soft masses. Upcoming dedicated stop and sbottom searches by ATLAS and CMS will potentially allow to test this particular corner of the parameter space.

The composition of the neutralino LSP depends strongly on the localisation properties of $\tau_R$. We typically find a higgsino-like LSP in the models with $\tau_R$ in the untwisted sector and a bino-like LSP in the models with $\tau_R$ in the twisted sector.
We have shown that for a higgsino LSP the correct dark matter density can be obtained non-thermally by dilaton decay, while in the bino case it can be produced thermally through stop coannihilations. Still, a large fraction of the parameter space with a bino LSP is disfavoured by thermal overproduction unless one allows for very special cosmological histories. 

If the neutralino LSP accounts for the observed dark matter, the direct detection cross section is generically large enough to be probed by the next generation direct detection experiments for the models with $\tau_R$ in the untwisted sector, while those with $\tau_R$ in the twisted sector would partially escape detection.

\section*{Acknowledgments}
In particular we would like to thank Michael Ratz for extended conversations at early stages of this project. Furthermore it is a pleasure to thank Ben Allanach and Jamie Tattersall for useful conversations. 
This work was partially supported by the SFB--Transregio TR33 ``The Dark Universe'' (Deutsche Forschungsgemeinschaft) and the European Union 7th network program ``Unification in the LHC era'' (PITN--GA--2009--237920). MB also acknowledges partial support by the National Science Centre in Poland under research grant DEC-2011/01/M/ST2/02466.

\appendix

\section{Determining $\varrho$ in heterotic orbifold compactifications}
\label{sec:detrho}

\begin{figure}[h]
\centering
  \includegraphics[height=4.4cm]{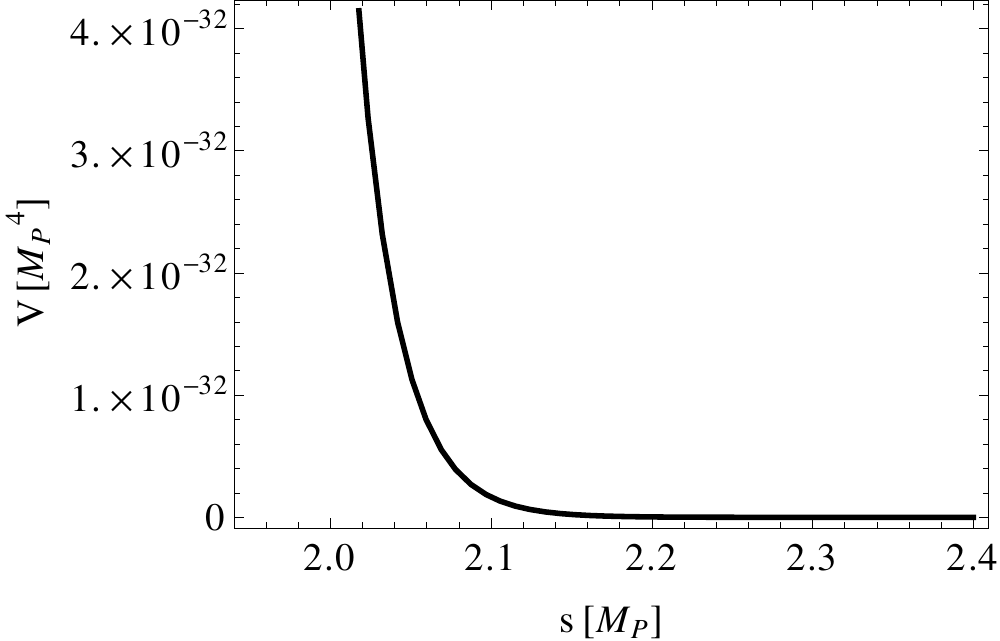}\hspace{5mm}
  \includegraphics[height=4.4cm]{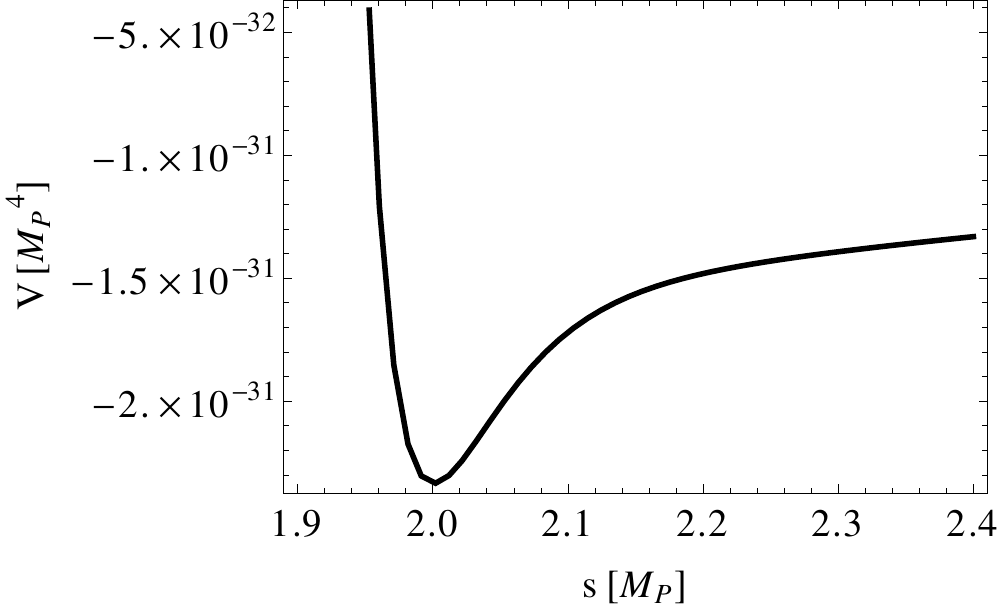}\caption{Run-away potential using the parameters $N=4$ and $P=1$ in equation~\eqref{eq:runawaypot} (left panel). The AdS potential for $N=4$ and $P=1$ with C adjusted using the relation from equation~\eqref{eq:crel} (right panel).}
\label{fig:runaway}
\end{figure}

Let us review in this appendix the scheme of moduli stabilisation for the heterotic orbifold compactifications in some detail, discussing an explicit example for a hidden matter sector that can be used for uplifting. Here we are interested in a setup that stabilises complex structure and K\"ahler moduli supersymmetrically, leaving only the dilaton $S$ unstabilised. The superpotential and leading order K\"ahler potential dependence of the dilaton are given by
\begin{eqnarray}
K&=&-\log{(S+\bar{S})}\, , \\
W&=& P e^{-b S}\, ,
\end{eqnarray}
where the coefficient $b$ in the exponent is determined by the rank of the gauge group inducing the hidden sector gaugino condensate; for $SU(N)$ $b$ is given by $b=8\pi^2/N.$ To obtain the correct gauge coupling at the unification scale, we need to stabilise the dilaton at $s_0 \simeq 2\, .$ The potential around this desired minimum is given by the typical run-away potential for the dilaton (also shown in figure~\ref{fig:runaway} on the left panel)
\begin{equation}
V=2 P^2 b^2 s e^{-2bs}\, ,
\label{eq:runawaypot}
\end{equation}
where $s$ is the real part of the dilaton. Note that we are in the limit $bs\gg 1$ for any reasonable value of the hidden sector gauge group.

To stabilise the dilaton, it is sufficient to include a constant contribution to the superpotential as the flux contribution in the type IIB context
\begin{equation}
W=C+P e^{-b S}\, .
\end{equation}
Such a hierarchically small constant can appear in the presence of approximate R-symmetries as discussed in~\cite{Kappl:2008ie}.
This allows to construct a supersymmetric minimum as in the type IIB framework of KKLT~\cite{Kachru:2003aw} by demanding a supersymmetric stabilisation of the dilaton $D_S W=0,$ leading to the following condition
\begin{equation}
-P e^{-b s} (1+2 b s)= C\, .
\label{eq:crel}
\end{equation}
Numerically this requires a hierarchically small constant $C\simeq 5.72\times 10^{-16}$ for $A=1$ and $N=4$ to obtain a minimum at $s=2.$ Applying this relation for $C$ for any value of $s$ leads to an AdS minimum (as shown in figure~\ref{fig:runaway} in the right panel) with a vacuum energy
\begin{equation}
V_0=-\frac{3|W|^2}{2 s}= - \frac{3|C+P e^{-2b}|^2}{4}=-3|Pbs|^2 e^{-2bs}\, .
\end{equation}
The gravitino mass is given by
\begin{equation}
m_{3/2}=\frac{|W|}{\sqrt{2s}}=\sqrt{2} P b e^{- b s_0} s_0=2P b e^{-2b}\, .
\label{eq:m32}
\end{equation}
To uplift the vacuum energy a positive contribution to the potential of the following type needs to be added
\begin{equation}
V_{\rm up}=\frac{r}{(2s)^m}\, ,
\end{equation}
where $m$ depends on the choice of the uplifting mechanism and
\begin{equation}
r=(2s_0)^m |V_0|\, .
\end{equation}
Depending on $m$ the minimum is slightly shifted through uplifting. This shift in $s,$ denoted by $\Delta s,$  can be determined by looking at the vanishing of the first derivative of the potential to be given by
\begin{equation}
\Delta s=\frac{3 m s_0}{-1+b s_0+2 b^2 s_0^2-3 m (1+2 b s_0)}\, .
\label{eq:ds}
\end{equation}
We are interested in the case $s_0=2$ and $m=1$ for which the shift in $s$ becomes\footnote{In KKLT constructions the uplifting usually scales as $m=3$ and hence leads to
$$\Delta \tau\simeq\frac{9}{2 b^2 \tau}\, ,$$ where $\tau$ represents the real part of the K\"ahler modulus $T$ that is stabilised by non-perturbative effects. 
}
\begin{equation}
\Delta s=-\frac{3}{2+5 b-4 b^2}\simeq \frac{3}{4 b^2}\, .
\end{equation}
Given this potential for the dilaton, one can in principle use various mechanisms to uplift the vacuum energy with F-terms or D-terms as discussed in the type IIB context. In realistic models from heterotic orbifold compactifications such a hidden sector for uplifting needs to be generated, but this is beyond the scope of this article and we restrict ourselves to a field theoretical toy-model at this stage. Here we discuss the example of a quantum corrected O'Raifeartaigh model as in~\cite{Kallosh:2006dv}, which allows for a minimum near the origin of the additional matter field $X.$ Such a matter construction with a minimum near the origin is desired as it guarantees the absence (respectively suppresses hierarchically) off-diagonal entries in the K\"ahler metric between the uplifting sector and the dilaton. If such an off-diagonal coupling is not sufficiently suppressed, additional contributions to the potential are generated and the original stabilisation procedure is changed. We take the following potential for the additional field $X$
\begin{equation}
W_X=-\mu^2 X\, , \qquad K_X= X \bar{X}-\frac{(X\bar{X})^2}{\Lambda^2}\, .
\end{equation}
We are interested in the limit (more justification on this limit can be found in~\cite{Kallosh:2006dv})
\begin{equation}
\mu^2,\Lambda^2\ll 1\, .
\end{equation}
Near $x\simeq 0$ and using the above relations for the coefficients, the potential is approximated by
\begin{equation}
V_X=\mu^4\left(1+\frac{x^2}{\Lambda^2}\right)\, .
\end{equation}
The potential clearly has a positive minimum around $x\simeq 0$ and hence can be used as an uplifting potential for the dilaton field.

Let us now discuss this combined potential of $X$ and $S$
\begin{eqnarray}
K&=& -\log{(S+\bar{S})}+X \bar{X}-\frac{(X\bar{X})^2}{\Lambda^2}\, ,\\
W&=& C + P e^{-b S}-\mu^2 X\, .
\end{eqnarray}
The imaginary parts are stabilised as before at zero, given that $C$ is negative which we assume from now on, otherwise the phases adjust such that this situation applies (see~\cite{Kallosh:2006dv} for more details). Focusing on the real parts, the potential can be written as
\begin{equation}
V=|F_S|^2+|F_X|^2-3 m_{3/2}^2\, ,
\end{equation}
where the gravitino mass is the one from equation~\eqref{eq:m32}. To cancel the vacuum energy $\mu$ is adjusted such that
\begin{equation}
F_X=\frac{\mu^2}{2}\simeq\sqrt{3}m_{3/2}=2 \sqrt{3}  P b e^{-2b}\, ,
\end{equation}
where the approximation is due to the slight shift in $X$ and $S$ and the resulting shift in their F-terms. Due to the small shift in $s,$ its F-term is no longer vanishing and we find to leading order
\begin{equation}
F_S=e^{K/2}K^{SS}D_SW\simeq 6P e^{-2b}=\frac{3 m_{3/2}}{b}\, .
\end{equation}
Given this structure in the F-terms on finds, as discussed in section~\ref{sec:scalesinminilandscape}, various phenomenologically interesting properties of the soft-masses, such as for example the suppression of the gaugino masses compared to the scalar masses.

\bibliography{heterotichiggsbib}
\bibliographystyle{h-physrev}

\end{document}